\documentclass[a4paper,fleqn,usenatbib,useAMS]{mnras}
\usepackage{newtxtext,newtxmath}
\usepackage[T1]{fontenc}
\usepackage{ae,aecompl}
\usepackage{graphicx}    
\usepackage{amsmath}     
\usepackage{multicol}    
\usepackage{bm}          
\usepackage{pdflscape}   
\usepackage{xspace}
\newcommand{\msun}{\,$M_{\sun}$}

\newcommand{\ergs}{\,erg\,s$^{-1}$}
\newcommand{\kms}{\,km\,s$^{-1}$}

\newcommand{\cmg}{\,cm$^2$\,g$^{-1}$}
\newcommand{\gcmq}{\,g\,cm$^{-3}$}
\newcommand{\gcm}{\,g\,cm$^{-1}$}

\newcommand{\nii}{\,[N\,II]\,6584\,\AA}
\newcommand{\ha}{\,H$\alpha$}
\newcommand{\hb}{\,H$\beta$}
\title[Circumstellar shell in SN 2008fq] 
  {Confined massive circumstellar shell in type IIL SN 2008fq}
\author[N. N. Chugai]{
N. N. Chugai\thanks{E-mail: nchugai@inasan.ru}
\\
$^{1}$Institute of Astronomy, Russian Academy of Sciences, Pyatnitskaya
      St. 48, 119017 Moscow, Russia\\
}

\date{Accepted XXX. Received YYY; in original form ZZZ}

\pubyear{2021}

\begin{document}
\label{firstpage}
\pagerange{\pageref{firstpage}--\pageref{lastpage}}
\maketitle
%
\begin{abstract}
I explore a phenomenon of the circumstellar \ha\ and \hb\ absorption lines 
 in the spectrum of 
 luminous type IIL SN~2008fq taken on day 6.8 after the discovery.
The absorpion is identified with the radiatively accelerated preshock wind up to $\sim800$\kms.
The required initial luminosity is attributed to the earlier circumstellar interaction with the 
  confined dense shell of $\sim0.08$\msun.
The modelling of similar luminous type IIL SN~1998S based on the same approach results in the 
  comparable shell mass of $\sim0.1$\msun. 
More than 1\,dex  larger mass of the confined dense shell of both 
  SNe~IIL compared to that of type IIP supernovae is attributed to the 
  larger progenitor mass of type IIL supernovae.
\end{abstract}

\section{Introduction} 
\label{sec:intro}

Ultimate understanding core collapse supernovae (CCSN) faces two major long-standing 
  problems: the poorly known explosion  mechanism and the relation between a certain 
  class of CCSN and a progenitor star. 
While the first problem is primarily a matter of the collapse and explosion theory, 
  the second is related to the theory of stellar evolution and the interpretation of 
  photometric and spectral observations of CCSN in terms of the progenitor mass.
The latter direction includes a growing significance of early spectra of CCSN that 
  reveal in many cases of type II supernovae signature of confined ($\sim 10^{15}$\,cm) 
  dense circumstellar (CS) shell \citep{Chugai_2001,Quimby_2007,Groh_2014,Khazov_2016,Yaron_2017}.
This phenomenon suggests that several years prior to 
   the explosion the mass loss is particularly vigorous that is related probably  
   to the nuclear burning in a mantle of the precollapse core.
 \citet{Shiode_2014} have argued that during and after core neon burning, internal  
  gravity waves excited by core convection can transport the energy of 
  nuclear burning out to the stellar envelope causing a robust mass loss. 
     
 The study of the confined dense CS shell in SNe~IIP might shed light on the 
  final stage of preSN~II of different mass that is crucial for the theory
  of CCSN.
 Remarkably, in the case of luminous SN~1998S (type IIL) with early CS emissions 
   the confined CS shell is rather massive, $\sim 0.1$\msun\ \citep{Chugai_2001}, 
   whereas in  SN~IIP SN~2013fs with the well 
   studied early CS emission lines the mass of the confined dense shell is 
   only $\sim 0.003$\msun\ \citep{Yaron_2017}. 
 The origin of this 30-fold difference is intriguing and may become a clue to the 
   physics of precollapse heavy mass loss of progenitor varieties. 
 Noteworthy, we atribute SN~1998S to SNe~IIL despite in some 
   papers it is referred to as SN~IIn.
 The presence of transient CS emission lines in early IIL and IIP supernovae obviously  
   is not a dominant typological property of these supernovae.
      
Luminous type II SN~2008fq was discovered 2008 March 18.4 \citep{Thrasher_2008}
  several days after the explosion \citep{Taddia_2013}.
Authors emphasise the strong similarity between SN~2008fq and SN~1998S, 
  although the early stage of strong CS emission lines has been missing in the former. 
On day 6.8 after discovery the spectrum shows a blue-shifted \ha\ and \hb\ absorption 
  with the width of $\sim700-800$\kms\ \citep{Taddia_2013}.
These absorptions undoubtedly originate from a dense CS gas; a broad \ha\ emission 
  with the  blue width at zero intensity (BWZI) of 7500\kms\ appears later, on day 29 \citep{Taddia_2013}. 
The CS \ha\ absorption of SN~2008fq brings to mind a shallow CS \ha\ absorption with 
  velocities in the blue wing up to $-500$\kms\ in early spectra of SN~1998S, in which  
  case this feature is accompanied by a deep narrow absorption formed in the slow wind 
  with the velocity  of $v_w = 40$\kms\ \citep{Fassia_2001}.
 The low resolution of SN~2008fq spectra ($\sim400$\kms) prevents one from the detection 
   of the similar slow wind.
The fast CS gas revealed by the \ha\ absorption in early spectrum of SN~2008fq presumably 
  has the same origin as that of SN~1998S, viz., a preshock CS gas accelerated 
   by the SN radiation \citep{Chugai_2002}. 
   
The velocity of the radiatively accelerated preshock wind has an important  
   diagnostic significance, since it is a measure of the radiation energy emitted between 
   the shock breakout and the moment of the spectral observation.
An interesting outcome of this reality is that the radiation energy 
  responsible for the CS preshock acceleration could be powered by the ejecta interaction 
  with a dense CS shell. 
This prompts us a possibility to employ the CS interaction model in order to 
  probe the confined CS shell. 
If success, SN~2008fq would become the second SN~1998S-like supernova with the 
  detected confined CS shell.
   
The proposed goal will be realized here in three steps. 
In Section \ref{sec:gen} we present a general picture of relevant phenomena. 
The \ha\ and \hb\ modelling in Section \ref{sec:abs} will provide us with the velocity of  
  the preshock gas in SN~2008fq. 
 We then use the CS interaction model to account for the preshock 
  velocity, bolometric light curve 
  and the late time ejecta velocity of SN~2008fq (Section \ref{sec:dyn}). 
The same modelling will be applied to SN~1998S in order to compare these events in 
 a uniform way.

The study is based on the SN~2008fq spectra \citep{Taddia_2013} retrieved from the 
  database WISeREP \citep{Yaron_2012}.  
The explosion was assumed to occur 4.5\,d before the discovery \citep{Taddia_2013} based on 
   the last non-dection of 9 days before the discovery.

 \section{General picture} 
\label{sec:gen}
   
\begin{figure}
 	\includegraphics[trim= 30 60 0 0,width=0.9\columnwidth]{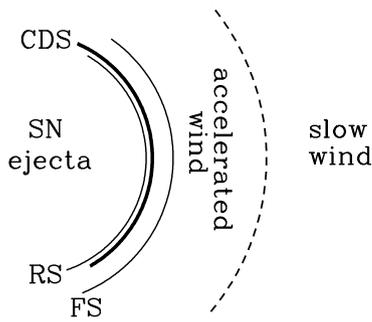}
 	\caption{%
 	Schematic representation of the model that accounts for the 
 	  early SN~2008fq spectrum.
 	Shown are opaque cold dense shell (CDS) that forms between forward shock (FS) and 
 	reverse shock (RS). On day 6.8 the photosphere coincides with the CDS.
 	Ahead of the FS shown are the radiatively accelerated wind  with the maximal velocity at the forward shock 
 	and external slow wind; both wind zones are responsible for the early CS \ha. 
 	}
 	\label{fig:cart}
 \end{figure}

First three weeks the SN~2008fq spectrum is featureless with low contrast \ha\ and \hb\ 
 narrow 
 ($-700...\,-800$\kms) absorption being most apparent detail apart from superimosed H\,II 
 region emission lines \citep{Taddia_2013}. 
The sketch (Fig. \ref{fig:cart}) illustrates how this spectrum presumably arises. 
The opaque thin cold dense shell (CDS) formed between forward and reverse shocks 
  blocks the radiation from SN ejecta; the similar situation is met in SN~1998S   \citep{Chugai_2001}. 
Outside the CDS there is the forward shock and preshock CS wind accelerated 
  by the radiation: this is the site where the \ha\ absorption 
  on day 6.8 does form.
The slow wind of the red supergiant (RSG) preSN lies further out.
The reliable value of the preshock velocity will be inferred in the next section  
   based on the modelling of the observed \ha\ and \hb\ avsorptions in the SN~2008fq. 
  
 %
 %
 
 The preshock acceleration at the radius $r$ due to the radiative force is 
    $a_r = kL_r/(4\pi r^2 c)$, where $k$ is the opacity dominated by the Thomson  
   scattering, $L_r$ is the radiative supernova luminosity,  
   and $c$ is speed of light. 
 The contribution of the line opacity into the radiative force with respect to 
  the Thomson scattering  
 for the CS density of $2\times10^{-15}$\gcmq\ in the case of 
  SN~1998S turns out to be   $\zeta \sim 0.3$ \citep{Chugai_2002}.
 This parameter gets lower for the larger wind density and shows a weak sensitivity to 
   the wind temperature in the range of $(1-4)\times10^4$\,K  \citep{Chugai_2002}.   
For $\zeta = 0.3$ and the solar composition the  
   effective opacity  is $k = 0.34(1 + \zeta) = 0.44$\cmg.
Neglecting the CS gas displacement the integration of the equation of motion 
  $dv/dt = a_r$ results in the preshock velocity
 \begin{equation}
   v_{ps} = \frac{kE_r}{4\pi r^2c} = 1170E_{r,50}r_{15}^{-2}\quad \mbox{\kms}\,,  
 \label{eq:accel}  
 \end{equation}  
where $E_{r,50}$ is the total radiation energy in units of $10^{50}$\,erg 
  emitted from the shock breakout till the certain age, $r_{15}$ is the radius in units 
  of $10^{15}$\,cm.    
On day 6.8 the photosphere radius is $\approx10^{15}$\,cm  \citep{Taddia_2013}, so 
  the preshock velocity of $\sim800$\kms\ for $r = 10^{15}$ 
  requires $E_r \sim7\times10^{49}$\,erg.
  
The required radiation energy is too large compared the expected for 
 the exploding RSG of well studied SN~IIP. 
In the case of the luminous type IIP SN~2004et with the explosion energy of  
  $2.3\times10^{51}$\,erg the energy radiated during first 10 days is only 
  $2\times10^{49}$\,erg \citep{UC_2009}.
This suggests that 
  the additional luminosity related to the CS interaction at the early stage 
 of SN~2008fq should be dominant.
Given the radical consequences for the early luminosity suggested by  $v_{ps}$ value,  
 the CS interaction model (Section \ref{sec:dyn})  will be constrained also by the observational 
 bolometric luminosity  and the maximal ejecta velocity at the later stage.

 %
 \begin{figure*}
 	\includegraphics[trim= 50 120 0 230,width=0.95\textwidth]{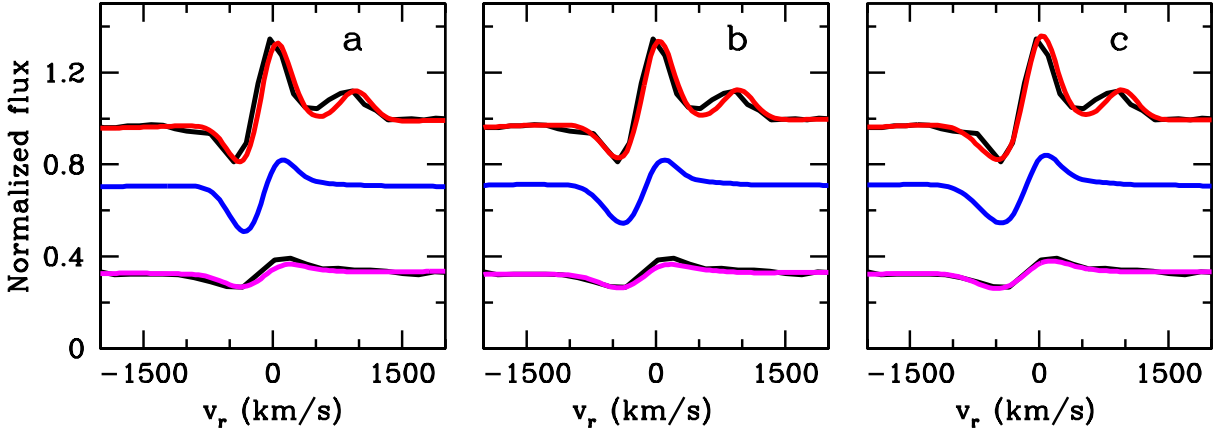}
 	\caption{%
 	 Model \ha\ with H\,II region lines ({\em up, red})
 	 and \hb\ spectra ({\em bottom, megenta}) 	 
 	 for the preshock velocity 600\kms\ (panel {\bf a}),
 	  800\kms\ (panel {\bf b}), and 1000\kms\ (panel {\bf c}) overploted 
 	  on the observed spectra ({\em black}). 
 	 Middle line ({\em blue}) shows \ha\ model without H\,II region lines of \ha\ and \nii.
 	 The case of 800\kms\ obviously is preferred for both \ha\ and \hb.
    	}
 	\label{fig:hamod}
 \end{figure*}
 %

 \section{Preshock velocity from \ha\ and \hb\ absorptions} 
\label{sec:abs}

The model for the \ha\ formation is a specified version of the cartoon in Figure \ref{fig:cart}.
The photosphere that coincides with the CDS ($r_p = r_{cds}$) is  
  encircled by the hot postshock layer with the forward shock radius $r_s$. 
The expansion of SN ejecta with a homologous kinematics ($v = r/t$) and the density 
  $\rho \propto v^{-8}$  occurs self-similarly in a steady wind ($\rho \propto 1/r^2$)   
    with $r_s/r_{cds} = 1.27$ \citep{Chevalier_1982s}. 
At the considered phase the forward shock is partially radiative, so the    
    postshock layer should be thiner; we adopt $r_s/r_{cds} = 1.2$.   
Fortunately the \ha\ absorption profile is not sensitive to this ratio.    
The radiatively accelerated preshock wind expands with the kinematics 
\begin{equation}
v = (v_{ps}-v_w)(r_s/r)^2 + v_w\,, 
\label{eq:kin}
\end{equation}
in line with the equation (\ref{eq:accel}), where $v_w$ is the velocity of the unperturbed wind 
  adopted to be 40\kms\ following SN~1998S.

The radiation transfer is calculated based on the Monte Carlo technique that takes into account continuum scattering in the CS gas,
     recombination \ha\ emissivity, resonant scattering, and Thomson scattering on thermal electrons with the angle-averaged frequency redistribution function \citep{Hummer_1967}. 
The line photon scattering at a certain resonant point is treated in the Sobolev  
  approximation.
 Yet the overall line photon scattering in the accelerated wind is essentially 
  a non-local process, in contrast to the Sobolev approximation.
 Indeed, we take into account that for the kinematics of equation (\ref{eq:kin})  
  there can be two resonant points along some rays. 
 Besides, the photon scattering off electrons with the thermal frequency redistribution 
  can result in the additional resonant scattering at different point, if the scattered photon acquires a  blueshift in the comoving frame.
     
The wind electron temperature is assumed to be constant and equal to the photospheric 
   temperature. 
The hydrogen ionization and the 2-nd level excitation is calculated using nebular  
  Saha-Boltzman equation. 
For the \hb\ line only the scattered emission is taken into account, whereas for the \ha\ 
  the net emissivity (recombination Case C) is included.
This is done to check weather the electron number density and the related Thomson 
 optical depth are consistent with the CS interaction model.
In fact, as will be shown, the net \ha\ emissivity turns out to be negligible 
  compared to the scattered continuum radiation.
   
The adopted wind density parameter $w = \dot{M}/v_w  = 4\times10^{16}$\,g\,cm$^{-1}$ 
  is implied by the interaction model (Section \ref{sec:dyn}); in this case the Thomson 
  optical depth outside the forward shock $\tau_{\mbox{\tiny T}} \approx 1$ on day 6.8.
The line photon can scatter in the hot postshock layer that is assumed to be uniform with 
  the density being four times of the  preshock density. 
Photons scattered on hot electrons are asumed to be lost since a scattered photon either 
  strikes the absorbing photosphere or 
  escapes the layer in the far wings due to the high electron thermal velocity of $\sim10^5$\kms.
  
The model \ha\ and \hb\ profiles are shown for cases $v_{ps}$ equal to 600\kms, 800\kms, 
  and 1000\kms\ (Figure \ref{fig:hamod}).
The computed spectrum is convolved with the Gaussian FWHM = 9.2\AA\ in order to 
  take into account the spectral resolution estimated from  H\,II region 
 \ha\ and [N\,II] 6584\,\AA\ lines. 
The nebular lines  with  Gaussian profiles and  
  the ratio \ha/[N\,II] = 2 indicated by the spectrum on day 71 are superimposed on the computed \ha.
Remarkably, the adopted line ratio is consistent with the median value for 
 the sample of giant H\,II regions in the ordinary spiral galaxy \citep{Briere_2012}.
The case of  $v_{ps} = 800$\kms\ is apparently preferred (Fig. \ref{fig:hamod}).
Noteworthy, the \ha\ is dominated by the scattering of the continuum radiation 
with the negligible contribution of the recombination emission.   
Interestingly, despite the simplified treatment of ionization and excitation, 
  the required electron temperature 12800\,K is comparable to the potospheric value 
  of $\sim 10000$\,K estimated from the blackbody continuum fit \citep{Taddia_2013}.

 I do not consider in detail an alternative possibility that  
   an energetic dynamic process could eject $\sim0.1$\msun\ with high velocities,
  up to 800\kms.  
 This option was already discussed in the case of SN~1998S \citep{Chugai_2002} and  
   was abandoned.
The point is that the violent ejection results in the ejecta kinematics of 
 $v \propto r$, in which case one expects an increasing absorption 
  velocity with time, in contrast to the observed velocity decrease \citep{Fassia_2001}.
Unfortunately, in the case of SN~2008fq the evolution of the absorption 
  velocity in \ha\ and \hb\ cannot be followed based on the available spectra 
   \citep{Taddia_2013}, so 
 formally one cannot rule out the energetic ejection.
The scenario of the accelerated slow wind is accepted here based on the strong 
 similarity between SN~2008fq and SN~1998S emphasised by \citet{Taddia_2013}.

\section{CS interaction and preshock acceleration}
\label{sec:dyn}

A question arises, what the early luminosity of SN~2008fq should be to 
  account for the preshock CS velocity of 800\kms\ on day 6.8 after the 
  discovery?
The order of magnitude estimate (Section \ref{sec:abs}) indicates that the 
   radiated energy at the initial stage 
  should be very large in order to provide the required preshock acceleration. 
I propose therefore that the early luminosity is powered by the CS interaction.

The CS interaction is treated based on the thin shell hydrodynamic approximation   
  \citep{Chevalier_1982x}, in which ejecta and the CS gas shocked 
  in the reverse and forward shocks, respectively, form a thin shell.  
Equations of motion and mass conservation are solved numerically.
Following approximation adopted earlier \citep{Chugai_2018} 
  the luminosity of the reverse and forward shock at the age $t$ is calculated as 
   the shock kinetic luminosity $L_k = 2\pi r_s^2\rho v_s^3$ (where  
   $r_s$ is the forward shock radius, $\rho$ 
   is the preshock CS density and $v_s$ the forward shock speed)   
   multiplied by the radiation efficiency $\eta = t/(t+t_c)$,
   where the cooling time $t_c$ is calculated for the postshock density being 
   four times 
   of the preshock density and assuming electron-ion equilibration.
The optical luminosity is equal to the luminosity of both shocks corrected for the 
  escaping X-ray luminosity, whereas for the preshock acceleration we use  
  the total radiation of both shocks.

The model for the optical luminosity includes also the luminosity powered by the 
  explosion and the radioactive decay of  $^{56}$Ni. 
This component is calculated using the \citet{Arnett_1980} analytical description of the  
  light curve.
The adopted $^{56}$Ni mass is 0.17\msun, comparable to 0.15\msun\ inferred for 
 SN~1998S \citep{Fassia_2000}. The large amount of  $^{56}$Ni is supported by 
  the presence of broad absorption lines of Na\,I and Ca\,II infrared triplet 
  in the latest spectrum on day 71 \citep{Taddia_2013}. 
This argument is based on the fact that ejecta absorption lines 
 require an internal energy source 
 that serves as a background continuum radiation. 
In the presence of both the internal and external (CS interaction) 
  energy source, the former should be comparable to the CS interaction  
  luminosity for the broad absorption line to be pronounced.
This can be illustrated using an approximate expression for the relative depth of the 
 SN broad absorption  $A = F_{int}[1 - \exp{(-\tau_s)}]/(F_{cs} + F_{int})$, 
  where $F_{cs}$ is the monochromatic quasi-continuum flux powered by the CS 
  interaction, $F_{int}$ is the monochromatic quasi-continuum flux related to the 
  internal source, and $\tau_s$ is the Sobolev optical depth.
In the extreme case, when the $F_{int}/F_{cs} \ll 1$ (e.g., SNe IIn) the 
   SN spectrum does not show broad absorption lines.

The assumed density distribution of SN ejecta is 
 $\rho = \rho_0/(1 + (v/v_0)^q$ with $q = 8$
 and parameters $\rho_0$, $v_0$ set by the ejecta mass $M$ and kinetic energy $E$.
The model luminosity, CDS velocity, and the CDS radius are determined only by the 
   CS density and the SN density $\rho(v)$ in external layers ($v > v_0$).
The ejecta density distribution in outer layers is invariant for 
  the mass and energy obeyed the relation $E \propto M^{(q-5)/(q-3)}$ or  
  $E \propto M^{0.6}$ in the case of $q = 8$.
  
\begin{figure}
   \includegraphics[trim=50 120 0 160,width=0.95\columnwidth]{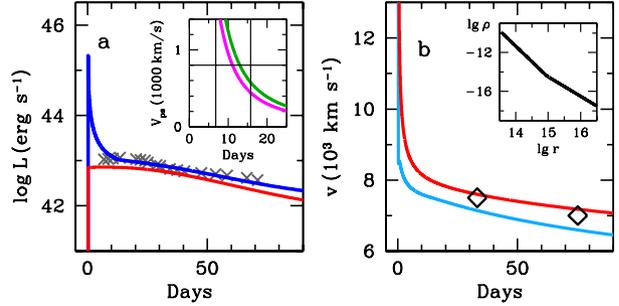}
   \caption{%
   The bolometric light curve and velocities in the model that includes the CS 
   interaction and $^{56}$Ni compared to observations.
   Panel {\bf a} shows the model bolometric light curve ({\em blue} line) overplotted 
   on the observational bolometric light curve ({\em crosses}). {\em Red} line is the 
   model light curve without the CS interaction. {\em Inset} shows the preshock velocity 
   for $\zeta = 0.3$ ({\em green}) and $\zeta = 0$ ({\em pink}).
   The horizontal line indicates the velocity 800\kms\ and vertical lines show the 
   the minimal and maximal time span since the explosion. 
   Panel {\bf b} shows the CDS velocity ({\em light blue}) and maximal velocity of unshocked 
   ejecta ({\em red}). {\em Diamonds} are the maximal velocity of the \ha\ line-emitting 
   gas from the BWZI velocity. {\em Inset} shows the density distribution of CS gas adopted in 
   the model.
    }
   \label{fig:dyn08fq}
\end{figure}

For the adopted ejecta mass of 10\msun\ the model that meets constraints from the 
  preshock velocity, bolometric light curve, and maximal velocity inferred from the 
  broad \ha\ emission  on days 29 and 71 suggests the kinetic energy 
   $E = 2\times10^{51}$\,erg and the wind density  
   $w = \dot{M}/v_w = 4\times10^{16}$\,g\,cm$^{-1}$ at $r > 9\times10^{14}$\,cm 
   (Fig. \ref{fig:dyn08fq}).
The inner zone $r < 9\times10^{14}$\,cm forms the CS shell with the mass 
   $M_{cs} = 0.08$\msun\ and the density $\rho \propto r^{-3.2}$.
It is the ejecta interaction with this shell that provides the most of the 
  radiation energy responsible for the radiative acceleration of the CS gas up to 800\kms\ at 
   10-13 days after the explosion.
This time span between the explosion and the spectral observation  
    (Fig. \ref{fig:dyn08fq}, panel {\bf a}, inset) is obtained for the accepted model 
   assuming the lines opacity contribution into the radiative force 
   $\zeta = 0$ and $\zeta = 0.3$.
The \ha\ recombination luminosity of the ionized wind on day 6.8 for the 
  temperature of 12000\,K is $\sim10^{39}$\ergs. 
This value is consistent with the negligible contribution of the 
  \ha\ net emission suggested by the \ha\ model (Section \ref{sec:abs}).
  
\begin{figure}
   \includegraphics[trim=50 120 0 160,width=0.95\columnwidth]{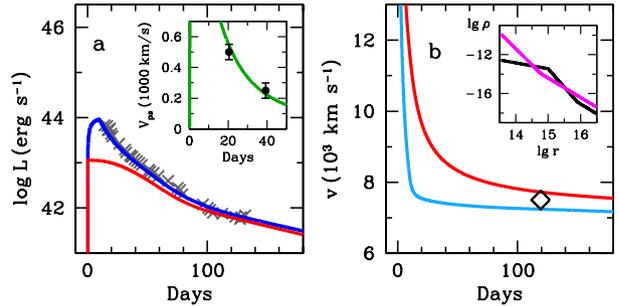}
   \caption{%
   The same as Figure \ref{fig:dyn08fq} but for SN~1998S.  
   Inset in the panel {\bf a} shows the preshock velocity compared to estimates from 
   blue wing of \ha\ absorption.
   ({\em Diamond}) in the panel {\bf b} is maximal velocity of the \ha\ line-emitting 
   gas from the BWZI velocity. {\em Inset}    
   shows the density of CS gas in the model of SN~1998S 
   ({\em black} line) compared to the CS density in the model of SN~2008fq.   
    }
   \label{fig:dyn98S}
\end{figure}

The similar modelling is performed for SN~1998S (Fig. \ref{fig:dyn98S}) with the 
  bolometric light curve from \citet{Fassia_2000}, and velocities from \citet{Fassia_2001}.
We adopt the ejecta mass of 10\msun\ and $^{56}$Ni mass of 0.15\msun.
The required kinetic energy $E = 3.2\times10^{51}$\,erg turns out to be a factor of 1.6 
  higher compared to SN~2008fq, while the  distant wind is more rarefied, 
  $w = 10^{16}$\,g\,cm$^{-1}$ at $r > 6\times10^{15}$\,cm.
This external wind density parameter coincides with the estimate based on the radio evolution 
  \citep{Pooley_2002}.
The confined CS shell mass at $r < 10^{15}$\,cm is $M_{cs} = 0.1$\msun\ that coincides 
  with the previous estimate \citep{Chugai_2001} recovered from effects of the Thomson 
  scattering on thermal electron and the light curve modelling.

%
%

%
%

  
\section{Discussion and Conclusions}
\label{sec:disc}

The study has been aimed at the extraction of the information on the close CS environment 
  of SN~2008fq based on the detected \ha\ and \hb\ CS absorption in early spectrum. 
We identify these absorpions with the radiatively accelerated preshock wind.
The required radiation for this acceleration is found to be powered by the ejecta interaction
  with the confined CS shell ($r <10^{15}$\,cm) containing 0.08\msun.
The similar modelling has been performed for the luminous type IIL SN~1998S in which case we 
  find the mass of the CS shell of 0.1\msun\ in line with the previous estimate 
  \citep{Chugai_2001}.

Remarkably, the confined CS shell in the well studied type IIP SN~2013fs \citep{Yaron_2017} 
contains only $\sim 0.003$\msun, more than 1\,dex lower compared to the CS shell mass 
 in both SNe~IIL.
The large difference indicates 
 that dynamic processes at the neon burning stage in SNe~IIL are more powerful 
 compared to SNe~IIP.
Note that the distant wind in both SNe~IIL ($w \gtrsim 10^{16}$\gcm) is also   
  significantly denser compared to the wind in SNe~IIP ($w < 10^{15}$\gcm) \citep{CFN_2006,CCU_2007}.  
A guiding parameter responsible for the mass loss difference 
  between SNe~IIL and SNe~IIP might be the progenitor mass.
  
Indeed, for the RSG mass loss driven by the radiation pressure   
  the energy conservation implies $(1/2)\dot{M}v_w^2 = \eta L(v_w/c)$, where $c$ is speed 
    of light, $M$ is assumed to be close to the initial progenitor mass, $\eta$ 
    is a fudge factor.
Since the wind velocity is of the order of the escape velocity $(2GM/R)^{0.5}$ 
  one gets then $w = \dot{M}/v_w = \eta LR/(GMc)$.
On the $L$ vs. $T_{eff}$ diagram the RSG stars occupy a narrow range    
   of temperatures around $T_{eff} = 4000$\,K \citep{Massey_2020}.
For the fixed $T_{eff}$ combined with the relation $L \propto M^{2.6}$ suggested by 
  the massive star evolutiom in the range of 15 - 25\msun\  \citep{Meynet_2015} 
 one obtains $w \propto M^{2.9}$.     
This supports the general view that the RSG wind density increases with the 
 progenitor mass. 

Recently \citep{Beasor_2020} inferred phenomenological relation between mass loss rate, 
  RSG  luminosity, and the initial star mass $\dot{M} \propto L^{4.8}\exp{(-0.23M)}$.
Using the evolutionary $L(M)$ relation and above assumptions concerning RSG values of 
  $v_w$ and $T_{eff}$
  one gets the relation $w \propto M^{12.6}\exp{(-0.23M)}$, which suggests the power 
  law $w \propto M^{\omega}$ with $\omega \sim 8...\,6$ in the mass range of 20 - 25\msun. 
  
Both $w(M)$ relations taken together with the wind parameter $w$ indicate that progenitors of 
 SNe~IIL are more massive compared to those of SNe~IIP.
This in turn is consistent  with larger $^{56}$Ni mass ($0.15-0.17$\msun) of both 
  SNe~IIL compared to the range of $0.01-0.08$\msun\ that is characteristic of 
  SNe~IIP \citep[cf.][]{UC_2019} and also is in line with the 
  correlation between $^{56}$Ni mass and the progenitor mass of CCSNe \citep{Hamuy_2003}.
Currently, SN~2008fq and SN~1998S are the only luminous SNe~IIL, 
  for which the $^{56}$Ni mass is estimated based on the modelling the bolometric light curve with both the CS interaction and the radioactive power taken into account. 
 The conjecture on the high $^{56}$Ni mass in luminous SNe~IIL, therefore, 
  requires further confirmation.
 
\section*{Acknowledgements}

I am grateful to Nando Patat for usefull discussions of early spectra of 
  SN~2008fq.

  
\section*{Data Availability}

Data available on request  
  

\bsp	

\begin{thebibliography}{}
\makeatletter
\relax
\def\mn@urlcharsother{\let\do\@makeother \do\$\do\&\do\#\do\^\do\_\do\%\do\~}
\def\mn@doi{\begingroup\mn@urlcharsother \@ifnextchar [ {\mn@doi@}
  {\mn@doi@[]}}
\def\mn@doi@[#1]#2{\def\@tempa{#1}\ifx\@tempa\@empty \href
  {http://dx.doi.org/#2} {doi:#2}\else \href {http://dx.doi.org/#2} {#1}\fi
  \endgroup}
\def\mn@eprint#1#2{\mn@eprint@#1:#2::\@nil}
\def\mn@eprint@arXiv#1{\href {http://arxiv.org/abs/#1} {{\tt arXiv:#1}}}
\def\mn@eprint@dblp#1{\href {http://dblp.uni-trier.de/rec/bibtex/#1.xml}
  {dblp:#1}}
\def\mn@eprint@#1:#2:#3:#4\@nil{\def\@tempa {#1}\def\@tempb {#2}\def\@tempc
  {#3}\ifx \@tempc \@empty \let \@tempc \@tempb \let \@tempb \@tempa \fi \ifx
  \@tempb \@empty \def\@tempb {arXiv}\fi \@ifundefined
  {mn@eprint@\@tempb}{\@tempb:\@tempc}{\expandafter \expandafter \csname
  mn@eprint@\@tempb\endcsname \expandafter{\@tempc}}}

\bibitem[\protect\citeauthoryear{{Arnett}}{{Arnett}}{1980}]{Arnett_1980}
{Arnett} W.~D.,  1980, \mn@doi [\apj] {10.1086/157898}, \href
  {https://ui.adsabs.harvard.edu/abs/1980ApJ...237..541A} {237, 541}

\bibitem[\protect\citeauthoryear{{Beasor}, {Davies}, {Smith}, {van Loon},
  {Gehrz}  \& {Figer}}{{Beasor} et~al.}{2020}]{Beasor_2020}
{Beasor} E.~R.,  {Davies} B.,  {Smith} N.,  {van Loon} J.~T.,  {Gehrz} R.~D.,
  {Figer} D.~F.,  2020, \mn@doi [\mnras] {10.1093/mnras/staa255}, \href
  {https://ui.adsabs.harvard.edu/abs/2020MNRAS.492.5994B} {492, 5994}

\bibitem[\protect\citeauthoryear{{Bri{\`e}re}, {Cantin}  \&
  {Spekkens}}{{Bri{\`e}re} et~al.}{2012}]{Briere_2012}
{Bri{\`e}re} {\'E}.,  {Cantin} S.,   {Spekkens} K.,  2012, \mn@doi [\mnras]
  {10.1111/j.1365-2966.2012.21450.x}, \href
  {https://ui.adsabs.harvard.edu/abs/2012MNRAS.425..261B} {425, 261}

\bibitem[\protect\citeauthoryear{{Chevalier}}{{Chevalier}}{1982a}]{Chevalier_1982s}
{Chevalier} R.~A.,  1982a, \mn@doi [\apj] {10.1086/160126}, \href
  {https://ui.adsabs.harvard.edu/abs/1982ApJ...258..790C} {258, 790}

\bibitem[\protect\citeauthoryear{{Chevalier}}{{Chevalier}}{1982b}]{Chevalier_1982x}
{Chevalier} R.~A.,  1982b, \mn@doi [\apj] {10.1086/160167}, \href
  {http://adsabs.harvard.edu/abs/1982ApJ...259..302C} {259, 302}

\bibitem[\protect\citeauthoryear{{Chevalier}, {Fransson}  \&
  {Nymark}}{{Chevalier} et~al.}{2006}]{CFN_2006}
{Chevalier} R.~A.,  {Fransson} C.,   {Nymark} T.~K.,  2006, \mn@doi [\apj]
  {10.1086/500528}, \href
  {https://ui.adsabs.harvard.edu/abs/2006ApJ...641.1029C} {641, 1029}

\bibitem[\protect\citeauthoryear{{Chugai}}{{Chugai}}{2001}]{Chugai_2001}
{Chugai} N.~N.,  2001, \mn@doi [\mnras] {10.1111/j.1365-2966.2001.04717.x},
  \href {https://ui.adsabs.harvard.edu/abs/2001MNRAS.326.1448C} {326, 1448}

\bibitem[\protect\citeauthoryear{{Chugai}}{{Chugai}}{2018}]{Chugai_2018}
{Chugai} N.~N.,  2018, \mn@doi [\mnras] {10.1093/mnras/sty2386}, \href
  {https://ui.adsabs.harvard.edu/abs/2018MNRAS.481.3643C} {481, 3643}

\bibitem[\protect\citeauthoryear{{Chugai}, {Blinnikov}, {Fassia}, {Lundqvist},
  {Meikle}  \& {Sorokina}}{{Chugai} et~al.}{2002}]{Chugai_2002}
{Chugai} N.~N.,  {Blinnikov} S.~I.,  {Fassia} A.,  {Lundqvist} P.,  {Meikle}
  W.~P.~S.,   {Sorokina} E.~I.,  2002, \mn@doi [\mnras]
  {10.1046/j.1365-8711.2002.05086.x}, \href
  {https://ui.adsabs.harvard.edu/abs/2002MNRAS.330..473C} {330, 473}

\bibitem[\protect\citeauthoryear{{Chugai}, {Chevalier}  \& {Utrobin}}{{Chugai}
  et~al.}{2007}]{CCU_2007}
{Chugai} N.~N.,  {Chevalier} R.~A.,   {Utrobin} V.~P.,  2007, \mn@doi [\apj]
  {10.1086/518160}, \href {http://adsabs.harvard.edu/abs/2007ApJ...662.1136C}
  {662, 1136}

\bibitem[\protect\citeauthoryear{{Fassia} et~al.,}{{Fassia}
  et~al.}{2000}]{Fassia_2000}
{Fassia} A.,  et~al., 2000, \mn@doi [\mnras]
  {10.1046/j.1365-8711.2000.03797.x}, 318, 1093

\bibitem[\protect\citeauthoryear{{Fassia} et~al.,}{{Fassia}
  et~al.}{2001}]{Fassia_2001}
{Fassia} A.,  et~al., 2001, \mn@doi [\mnras]
  {10.1046/j.1365-8711.2001.04282.x}, \href
  {https://ui.adsabs.harvard.edu/abs/2001MNRAS.325..907F} {325, 907}

\bibitem[\protect\citeauthoryear{{Groh}}{{Groh}}{2014}]{Groh_2014}
{Groh} J.~H.,  2014, \mn@doi [Astronomy and Astrophysics]
  {10.1051/0004-6361/201424852}, \href
  {https://ui.adsabs.harvard.edu/abs/2014A&A...572L..11G} {572, L11}

\bibitem[\protect\citeauthoryear{{Hamuy}}{{Hamuy}}{2003}]{Hamuy_2003}
{Hamuy} M.,  2003, \mn@doi [\apj] {10.1086/344689}, \href
  {https://ui.adsabs.harvard.edu/abs/2003ApJ...582..905H} {582, 905}

\bibitem[\protect\citeauthoryear{{Hummer} \& {Mihalas}}{{Hummer} \&
  {Mihalas}}{1967}]{Hummer_1967}
{Hummer} D.~G.,  {Mihalas} D.,  1967, \mn@doi [\apjl] {10.1086/180092}, \href
  {https://ui.adsabs.harvard.edu/abs/1967ApJ...150L..57H} {150, L57}

\bibitem[\protect\citeauthoryear{{Khazov} et~al.,}{{Khazov}
  et~al.}{2016}]{Khazov_2016}
{Khazov} D.,  et~al., 2016, \mn@doi [\apj] {10.3847/0004-637X/818/1/3}, \href
  {https://ui.adsabs.harvard.edu/abs/2016ApJ...818....3K} {818, 3}

\bibitem[\protect\citeauthoryear{{Massey}, {Neugent}, {Levesque}, {Drout}  \&
  {Courteau}}{{Massey} et~al.}{2020}]{Massey_2020}
{Massey} P.,  {Neugent} K.~F.,  {Levesque} E.~M.,  {Drout} M.~R.,   {Courteau}
  S.,  2020, arXiv e-prints, \href
  {https://ui.adsabs.harvard.edu/abs/2020arXiv201113279M} {p. arXiv:2011.13279}

\bibitem[\protect\citeauthoryear{{Meynet} et~al.,}{{Meynet}
  et~al.}{2015}]{Meynet_2015}
{Meynet} G.,  et~al., 2015, \mn@doi [\aap] {10.1051/0004-6361/201424671}, \href
  {https://ui.adsabs.harvard.edu/abs/2015A&A...575A..60M} {575, A60}

\bibitem[\protect\citeauthoryear{{Pooley} et~al.,}{{Pooley}
  et~al.}{2002}]{Pooley_2002}
{Pooley} D.,  et~al., 2002, \mn@doi [\apj] {10.1086/340346}, \href
  {https://ui.adsabs.harvard.edu/abs/2002ApJ...572..932P} {572, 932}

\bibitem[\protect\citeauthoryear{{Quimby}, {Wheeler}, {H{\"o}flich}, {Akerlof},
  {Brown}  \& {Rykoff}}{{Quimby} et~al.}{2007}]{Quimby_2007}
{Quimby} R.~M.,  {Wheeler} J.~C.,  {H{\"o}flich} P.,  {Akerlof} C.~W.,  {Brown}
  P.~J.,   {Rykoff} E.~S.,  2007, \mn@doi [\apj] {10.1086/520532}, \href
  {https://ui.adsabs.harvard.edu/abs/2007ApJ...666.1093Q} {666, 1093}

\bibitem[\protect\citeauthoryear{{Shiode} \& {Quataert}}{{Shiode} \&
  {Quataert}}{2014}]{Shiode_2014}
{Shiode} J.~H.,  {Quataert} E.,  2014, \mn@doi [\apj]
  {10.1088/0004-637X/780/1/96}, \href
  {https://ui.adsabs.harvard.edu/abs/2014ApJ...780...96S} {780, 96}

\bibitem[\protect\citeauthoryear{{Taddia} et~al.,}{{Taddia}
  et~al.}{2013}]{Taddia_2013}
{Taddia} F.,  et~al., 2013, \mn@doi [\aap] {10.1051/0004-6361/201321180}, 555,
  A10

\bibitem[\protect\citeauthoryear{{Thrasher}, {Li}  \& {Filippenko}}{{Thrasher}
  et~al.}{2008}]{Thrasher_2008}
{Thrasher} P.,  {Li} W.,   {Filippenko} A.~V.,  2008, Central Bureau Electronic
  Telegrams, \href {https://ui.adsabs.harvard.edu/abs/2008CBET.1507....1T}
  {1507, 1}

\bibitem[\protect\citeauthoryear{{Utrobin} \& {Chugai}}{{Utrobin} \&
  {Chugai}}{2009}]{UC_2009}
{Utrobin} V.~P.,  {Chugai} N.~N.,  2009, \mn@doi [\aap]
  {10.1051/0004-6361/200912273}, \href
  {https://ui.adsabs.harvard.edu/abs/2009A&A...506..829U} {506, 829}

\bibitem[\protect\citeauthoryear{{Utrobin} \& {Chugai}}{{Utrobin} \&
  {Chugai}}{2019}]{UC_2019}
{Utrobin} V.~P.,  {Chugai} N.~N.,  2019, \mn@doi [\mnras]
  {10.1093/mnras/stz2716}, \href
  {https://ui.adsabs.harvard.edu/abs/2019MNRAS.490.2042U} {490, 2042}

\bibitem[\protect\citeauthoryear{{Yaron} \& {Gal-Yam}}{{Yaron} \&
  {Gal-Yam}}{2012}]{Yaron_2012}
{Yaron} O.,  {Gal-Yam} A.,  2012, \mn@doi [\pasp] {10.1086/666656}, \href
  {https://ui.adsabs.harvard.edu/abs/2012PASP..124..668Y} {124, 668}

\bibitem[\protect\citeauthoryear{{Yaron} et~al.,}{{Yaron}
  et~al.}{2017}]{Yaron_2017}
{Yaron} O.,  et~al., 2017, \mn@doi [Nature Physics] {10.1038/nphys4025}, \href
  {https://ui.adsabs.harvard.edu/abs/2017NatPh..13..510Y} {13, 510}

\makeatother
\end{thebibliography}
\end{document}